\documentclass[12pt]{iopart}
\usepackage[british]{babel}
\usepackage[utf8]{inputenc}
\usepackage{amsmath, amssymb, bm}
\usepackage{graphicx}
\usepackage{appendix}
\usepackage{hyperref}
\usepackage{siunitx}

\renewcommand{\vec}[1]{\mathbf{#1}}

\begin{document}

\title[QM Thermo-Optics]{Hartree-Fock Analogue Theory\\of Thermo-Optic Interaction}

\author{Enrico Stein}
\ead{estein@rhrk.uni-kl.de}
\address{Department of Physics and Research Center OPTIMAS, Technische Universität Kaiserslautern, Erwin-Schrödinger Straße 46, 67663 Kaiserslautern, Germany}

\author{Axel Pelster}
\ead{axel.pelster@physik.uni-kl.de}
\address{Department of Physics and Research Center OPTIMAS, Technische Universität Kaiserslautern, Erwin-Schrödinger Straße 46, 67663 Kaiserslautern, Germany}

\vspace{10pt}
\begin{indented}
\item[]\today
\end{indented}

\begin{abstract}
Thermo-optic interaction significantly differs from the usual particle-particle interactions in physics, as it is retarded in time. A prominent platform for realising this kind of interaction are photon Bose-Einstein condensates, which are created in dye-filled microcavities. The dye solution continually absorbs and re-emits these photons, causing the photon gas to thermalise and to form a Bose-Einstein condensate. Because of a non-ideal quantum efficiency, these cycles heat the dye solution, creating a medium that provides an effective thermo-optic photon-photon interaction. So far, only a mean-field description of this process exists.\\
This paper goes beyond by working out a quantum mechanical description of the effective thermo-optic photon-photon interaction. To this end, the self-consistent modelling of the temperature diffusion builds the backbone of the modelling. Furthermore, the manyfold experimental timescales allow for deriving an approximate Hamiltonian. The resulting quantum theory is applied in the perturbative regime to both a harmonic and a box potential for investigating its prospect for precise measurements of the effective photon-photon interaction strength.
\end{abstract}

%
\vspace{2pc}
%
\noindent{\submitto{\NJP}}

\maketitle

\section{Introduction}
\begin{figure}
	\centering
	\includegraphics[width=\linewidth]{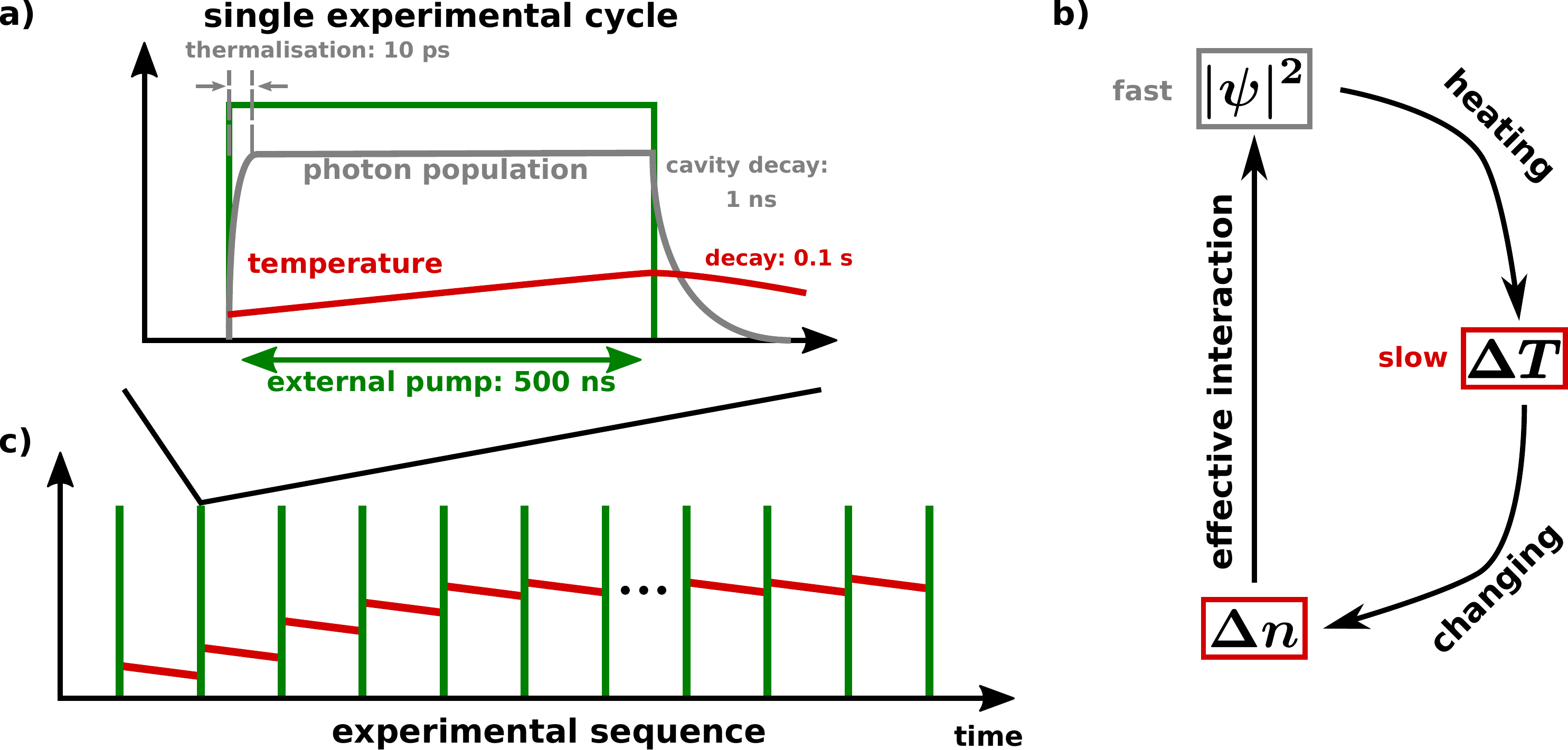}
	\caption{Timescales and interaction in typical photon BEC experiments, e.g., \cite{Klaers2010, Klaers2011}. \textbf{a)} Timescales in a single experimental cycle. The green line visualises the external pump pulse, the grey line the corresponding temporal evolution of the total photon population and the red line marks the behaviour of the temperature, which is produced by the photons. \textbf{b)} Emergence of the effective photon-photon interaction. Here $|\psi|^2$ stands for the photon density, $\Delta T$ for the temperature produced by the photons, and $\Delta n$ denotes the resulting shift of the refractive index. \textbf{c)} Temperature steady state after several experimental cycles. The green bars represent single experiments, like in \textbf{a)}, and the red line indicates the evolution of the temperature between two single experiments.}
	\label{Fig:Scales}
\end{figure}

Ultracold quantum gases usually deal with a contact particle interaction, since in these systems only s-wave scattering takes place due to the very low involved energy scales \cite{Pethick2008, Stringari2016}. Thus, two particles have to be at the same time at the same place for a scattering event to happen. Experiments with dipolar quantum gases loosen the latter restriction. Here, also particles at different places can interact with each other via the dipole-dipole interaction, which is both anisotropic and slowly decreasing in space \cite{Lahaye2009}. The scope of this work, however, lies on thermo-optic interactions, that are both non-local in space and retarded in time \cite{Boyd2020}. Therefore, two particles can interact with each other even though they are neither in proximity nor meet at the same time. Examples for thermo-optic non-linearities can be found in various settings. For instance, this kind of non-linearity plays a crucial role in analogue gravity, where it is used for simulating the Newton-Schrödinger equation \cite{Faccio2015, Bekenstein2015}. Moreover, thermo-optic effects are also used for tuning nanoresonators \cite{Tsoulos2020, Pruessner2007}. However, photon Bose-Einstein condensates (phBEC) provide another well controllable environment for observing this kind of unusual interaction. The theoretical description of thermo-optic interaction in phBECs is at the very focus of this work.\\
Photon Bose-Einstein condensates contain many competing timescales, which are schematically summarised in figure \ref{Fig:Scales}, a). A dye-filled microcavity is the main part of the experimental setup \cite{Klaers2010, Klaers2011}. The dye molecules set the fastest timescale by absorption and re-emission of photons ($\sim$ 1 ps). Since the vibrations of the dye molecules thermalise due to surrounding solvent molecules, the photon gas itself thermalises at the molecular timescale ($\sim$ 10 ps). Two more timescales determine the condensate lifetime. On the one hand, light leaks out of the cavity ($\sim$ 1 ns), and, on the other hand, the duration of counteracting external pump pulses is limited by dye bleaching ($\sim$ 500 ns). Finally, the heating of the whole experimental setup introduces the slowest timescale ($\sim$ 0.1 s). This heating stems from electronic excitations of the dye molecules, which are not remitted as photons, but are converted into vibronic excitations of the dye molecules. The temperature increase $\Delta T$, which the incoherent photon absorption processes produce, changes the refractive index of the dye solution by the amount $\Delta n \approx -\Delta T \partial n/\partial T$, ultimately leading to an effective thermo-optic photon-photon interaction, see figure \ref{Fig:Scales}, b) and references \cite{Klaers2010, Klaers2011, Stein2019} for details. Since this temperature diffuses through the cavity, the resulting effective photon-photon interaction is non-local in space and retarded in time.\\
Hitherto, 
former works have only focused on how the effective photon-photon interaction influences the phBEC ground state. For instance, the first publication on the experimental realisation of phBECs investigates, amongst other things, the effective photon-photon interaction by using a Gross-Pitaevskii equation for the modelling \cite{Klaers2010}. Introducing a coupled Schrödinger and temperature-diffusion equation improves the model on physical grounds \cite{Dung2017}. Here, the Schrödinger equation describes the fast evolution of the phBEC ground state, whereas the diffusion equation describes the slow dynamics of the temperature. The above-mentioned incoherent photon absorption processes steadily produce the latter. This model successfully describes the photon-photon interaction in the steady state of the temperature diffusion, which follows after several pump pulses, see figure \ref{Fig:Scales}, c). Further, this model allows for calculating the lowest-lying collective mode frequencies of the condensate \cite{Stein2019} as well as its intricate behaviour at the dimensional crossover from 2D to 1D \cite{Stein2022b}. The authors of reference \cite{Dung2017} succeed in describing the thermo-optic interaction emerging during a single pump pulse by applying a heuristic approximation, which relies on the photon timescales being much shorter than the temperature diffusion time.\\
Current measurements of the effective photon-photon interaction bear several disadvantages, as they are based on determining the condensate width. First, this approach relies on using the spatial data only, whilst the spectral data is, in principle, available at the same time. Utilising instead all experimentally accessible data promises to enhance the measurement accuracy for the photon-photon interaction strength. Second, these methods only consider the ground mode, which necessitates a large condensate fraction. But state-of-the-art experiments are only capable of achieving condensate fractions of about 50 \%, so the impact of the thermal cloud may not be neglected in a theoretical description. Third, the trapping geometry must be known well enough to obtain reliable information about the interaction-induced condensate broadening. To this end, the current experimental trend of realising more sophisticated trapping potentials is advantageous. Standard phBEC experiments use an isotropic harmonic potential \cite{Klaers2010, Klaers2011, Dung2017, Greveling2018} or double-well potentials \cite{Kurtscheid2019}. Reference \cite{Walker2018} reports the realisation of micro-condensates with only a few photons in such setups. Recently, experiments have even achieved box potentials for phBECs \cite{Busley2021}.\\
The theoretical description presented in this paper paves the way to more precise measurements of the effective photon-photon interaction strength, which are based on performing a detailed spectrometric analysis of the photon gas. This demands to extend the previous mean-field modelling by working out the underlying second-quantised Hamiltonian of the full photon field coupled to the temperature diffusion. A formal elimination of the temperature by its Green’s function leads to a compact expression for the resulting phBEC Hamiltonian during a single pump pulse, when taking the respective experimental timescales into account. Moreover, abstracting from single absorption/re-emission processes coarse grains the evolution during a single pump pulse and leads to a thermal photon gas in this description. Therefore, this procedure treating the thermo-optic interaction includes the thermal cloud self-consistently and corresponds to the usual Hartree-Fock approximation used, e.g., for atomic BECs in order to describe the impact of a contact interaction at finite temperature \cite{Pethick2008, Stringari2016}.\\
The paper is structured as follows: In section \ref{sec:Hamiltonian} the underlying second-quantised Hamiltonian describing the effective thermo-optic photon-photon interaction is introduced and simplified according to the respective experimental timescales. Finally, section \ref{sec:Perturbation} provides a perturbative calculation of the first few eigenenergies subject to the thermo-optic interaction and elucidates its perspective for precisely measuring the effective photon-photon interaction strength.
\section{Thermo-Optic Hamiltonian}\label{sec:Hamiltonian}
This section starts with formulating the basis of the quantum mechanical description of the thermo-optic interaction. To this end, the modelling considers the dynamics of both the second-quantised photon field and the temperature, that is produced by the incoherent photon absorption processes. Subsequently, taking the common experimental timescales into account provides a simplification of this general formulation with an approximate Hamiltonian.
\subsection{Generic Formulation}
The photon field operators $\hat\Psi(\vec x, t)$, $\hat\Psi^\dagger(\vec x, t)$ describe the electric field inside the cavity and fulfil the standard bosonic equal-time commutation relations,
\begin{align}
\left[\hat\Psi(\vec x, t), \hat\Psi(\vec x',t) \right] =\left[\hat\Psi^\dagger(\vec x, t), \hat\Psi^\dagger(\vec x',t) \right] =0 \, , \hspace*{0.5cm}
\left[\hat\Psi(\vec x, t), \hat\Psi^\dagger(\vec x',t) \right] = \delta(\vec x-\vec x')\, .
\end{align}
The corresponding second-quantised Hamiltonian includes the energy shift due to the temperature $\Delta T$, produced by the photons during the experiment, and reads
\begin{align}\label{eq:Ham}
\hat{H} (t) = \int d^2x~\hat\Psi^\dagger(\vec x, t)\Big\{ h (\vec x)+ \gamma \Delta T(\vec x, t)\Big\}\hat\Psi(\vec x, t)\,.		
\end{align}
Here,
\begin{align}\label{eq:h0}
h (\vec x)= -\frac{\hbar^2\bm\nabla^2}{2m} + V(\vec x)		
\end{align}
denotes the first-quantised Hamiltonian containing the effective photon mass $m$ and the trapping potential $V(\vec x)$, whereas the parameter $\gamma = -mc^2/n~\partial n/\partial T$ quantifies the energy shift due to the thermo-optic effect, see figure \ref{Fig:Scales}, b) and reference \cite{Stein2019}.\\ 
Conversely, the temperature $\Delta T(\vec x, t)$ obeys the diffusion equation
\begin{align}\label{eq:diff}
\frac{\partial \Delta T(\vec x, t)}{\partial t} = \left(\mathcal D\bm\nabla^2-\frac{1}{\tau}\right)\Delta T (\vec x, t)+ B n(\vec x, t)\,.		
\end{align}
Here, $\mathcal D$ denotes the diffusion coefficient of the solvent medium, $\tau$ the longitudinal relaxation time, and $B$ the heating coefficient of the dye solution \cite{Stein2019}. Furthermore, the photon density
\begin{align}\label{eq:density}
n (\vec x, t)= \left\langle\hat \Psi^\dagger(\vec x, t)\hat\Psi(\vec x, t)\right\rangle\, ,
\end{align}
with $\langle \bullet \rangle$ denoting the quantum mechanical expectation value, represents the source of the temperature $\Delta T(\vec x, t)$. Using the Green's function
\begin{align}\label{eq:Green}
\mathcal G (\vec x, t)= \frac{1}{4\pi \mathcal D t}\,e^{-\vec x^2/4\mathcal D t -t/\tau}
\end{align}
allows solving the diffusion equation \eqref{eq:diff} according to
\begin{align}\label{eq:temp}
\Delta T (\vec x, t)= \tau B \int_0^{t}\frac{dt'}{\tau}\int d^2x'~\mathcal G(\vec x-\vec x', t- t')n(\vec x', t') + \Delta T_0(\vec x) \,,
\end{align}
with the initial temperature distribution $\Delta T_0(\vec x)$, which may stem from prior pump pulses and accumulate during the experimental cycle, as is shown in figure \ref{Fig:Scales}, c) and discussed in \ref{app:steady_temp}. Inserting equation \eqref{eq:temp} into the second-quantised Hamiltonian \eqref{eq:Ham} yields
\begin{align}\label{eq:Ham1}
\begin{split}
\hat{H} (t) = \int d^2x~\hat\Psi^\dagger(\vec x, t)\Big\{& h (\vec x)+ g_T \int_0^{t}\frac{dt'}{\tau}\int d^2x'~\mathcal G(\vec x-\vec x', t- t')n(\vec x', t') \\
&+ \gamma\Delta T_0(\vec x)\Big\}\hat\Psi(\vec x, t)\,,		
\end{split}
\end{align}
with the steady-state interaction strength $g_T = \gamma\tau B$ \cite{Stein2019}. Note that the integral kernel in the Hamiltonian \eqref{eq:Ham1} also acts on the time, reflecting the temporal retardation of the thermo-optic interaction. This implies the condensate to not only depend on the current time, but also on its complete history. Reference \cite{Stein2019} investigates the impact of this memory effect upon the lowest-lying collective frequencies of a photon BEC subject to the thermo-optic interaction in the long-time lime. It turns out that the memory effect leads to a decrease of the collective frequencies, including the centre-of-mass oscillation. Figure \ref{Fig:Scales} a) clearly shows that the experiment can investigate the long-time limit only after several pump cycles. Therefore, the current work focuses on the short-time behaviour during a single pump pulse and calculates the appearing energy shifts, which are equivalent to the lowest-lying collective excitations for the long timescales. Therefore, the next section shows, how the Hamiltonian \eqref{eq:Ham1} can be simplified by considering the experimental timescales of a single pump pulse.

\subsection{Experimental Timescales}
Rescaling the integration variable $t'$ in equations \eqref{eq:temp} and subsequently in \eqref{eq:Ham1} by $1/\tau$ explicitly reveals the $t/\tau$ dependency of the integral. Experimentally, $t$ corresponds to the phBEC lifetime and $\tau$ denotes the decay time of the temperature difference, cf.~figure \ref{Fig:Scales}, a) and \ref{app:steady_temp}. Since these timescales imply the ratio $t/\tau\sim10^{-6}$, an expansion up to the first order in $t/\tau$ yields an accurate approximation for the temperature \eqref{eq:temp}. In particular, the Gaussian function in \eqref{eq:Green} goes over into a Dirac-$\delta$ distribution and \eqref{eq:temp} reduces to
\begin{align}\label{eq:temp_approx}
\Delta T(\vec x, t)\approx \tau B \, \frac{t}{\tau}\,n(\vec x, 0)		
\end{align}
As a consequence of this approximation, the details of the Green's function \eqref{eq:Green} are irrelevant to the further experiment, and only the initial photon density $n(\vec x,0)$ defined according to equation \eqref{eq:density} determines the spatial temperature profile. This is a direct outcome of the temporal retardation of the thermo-optic interaction. Therefore, this type of interaction prevents a back action of the actual photon density.\\
The same approximation that lead to the temperature distribution  \eqref{eq:temp_approx} yields for the second-quantised Hamiltonian \eqref{eq:Ham1}
\begin{align}\label{eq:Ham2}
\hat{H} (t)= \int d^2x~\hat\Psi^\dagger(\vec x, t)\Big\{ h (\vec x)+ g(t)n(\vec x, 0)\Big\}\hat\Psi(\vec x, t)\,, 	
\end{align}
with the effective time-dependent thermo-optic interaction strength
\begin{align}\label{eq:g}
g(t) = t\gamma B \,.		
\end{align}
Conclusively, the thermo-optic interaction behaves like an effective potential, that increases linearly in time, rather than like a usual two-particle interaction, which is local in time. This is the immediate consequence of the interplay between the slow growth of the temperature during a single pump pulse and the fast thermalisation timescale.%
\subsection{Adiabatic Treatment}
Since the temperature timescale is by far the slowest, as figure \ref{Fig:Scales}, a) illustrates, treating the time dependence of the interaction strength \eqref{eq:g} adiabatically is justified \cite{Sakurai}. The aim is to investigate the instantaneous steady states of the second-quantised Hamiltonian \eqref{eq:Ham2}. To this end, the eigenvalue problem of the first-quantised Hamiltonian \eqref{eq:h0}
\begin{align}
h(\vec x)\psi_{\bm n}(\vec x) = E_{\bm n}(0)\psi_{\bm n}(\vec x)
\end{align}
yields a basis of orthonormal eigenmodes $\psi_{\bm n}(\vec x)$ with corresponding eigenenergies $E_{\bm n}(0)$. Here, $\bm n$ is a multi-index denoting all the quantum numbers of the corresponding state. This provides an expansion of the field operators 
\begin{align}\label{eq:modes}
\hat\Psi (\vec x, t)=\sum_{\bm n} \hat a_{\bm n}(t) \psi_{\bm n}(\vec x)	\, , \hspace*{1cm}\hat\Psi^\dagger (\vec x, t)=\sum_{\bm n} \hat a_{\bm n}^\dagger(t) \psi_{\bm n}^*(\vec x)\, .	
\end{align}
Here, the annihilation and creation operators $\hat a_{\bm n}(t)$ and $\hat a_{\bm n}^\dagger(t)$ fulfil the canonical bosonic commutation relations:
\begin{align}
\Big[ \hat a_{\bm n}(t)  , \hat a_{\bm{n'}}(t)  \Big] =\Big[ \hat a_{\bm n}^\dagger(t)  , \hat a^\dagger_{\bm{n'}}(t) \Big] =0 \, , \hspace*{1cm}
\Big[ \hat a_{\bm n}(t)  , \hat a^\dagger_{\bm {n'}}(t)  \Big] = \delta_{{\bm n},{\bm{n'}}}\, .
\end{align}
As at the beginning of the experiment no interaction is present, the annihilation, and creation operators $\hat a_{\bm n}(0)$ and $\hat a_{\bm n}^\dagger(0)$ belong to the plain eigenmodes $\psi_{\bm n}(\vec x)$ of the first-quantised Hamiltonian \eqref{eq:h0}. Moreover, since the pump laser determines the polarisation of the photon field in the condensed phase \cite{Moodie2017, Greveling2017}, only a single photon polarisation is present, thus, the annihilation and creation operators do not carry a polarisation index. The expansion \eqref{eq:modes} allows writing the second-quantised Hamiltonian \eqref{eq:Ham2} in the form
\begin{align}
\hat H (t) = \sum_{\bm{n} \bm{n'}}\mathcal{H}_{\bm{n}, \bm{n'}} (t) \hat a^\dagger_{\bm n}(t) \hat a_{\bm{n'}}(t)\,,
\end{align}
with the Hamiltonian matrix
\begin{align}\label{eq:Hmat}
\mathcal H_{\bm n, \bm{n'}} (t)= E_{\bm n}{(0)} \delta_{\bm n, \bm{n'}} + g(t) F_{\bm n, \bm{n'}}\,.
\end{align}
The non-diagonal matrix
\begin{align}\label{eq:Fmat1}
F_{\bm n, \bm{n'}} = \int d^2x~\psi_{\bm n}^* (\vec x) n(\vec x, 0) \psi_{\bm{n'}}(\vec x)
\end{align}
contains the overlap of two modes $\bm n\,,\bm{n'}$ with the initial density $n(\vec x, 0)$ and, thus, describes the influence of the thermo-optic interaction. Diagonalising the Hamiltonian matrix \eqref{eq:Hmat} determines the finite-time operators $\hat a_{\bm n}(t)$, $\hat a_{\bm n}^\dagger(t)$, specifying the instantaneous eigenmodes. Thus, the latter take the form of a superposition of the discrete eigenmodes, determined by the external potential $V(\vec x)$, as time increases.
\subsection{Thermal Steady State}
In the following, this work does not include the actual thermalisation dynamics of the photons, but instead focuses on the long timescales during a single pump pulse, where the influence of the thermo-optic interaction  becomes of interest, cf., figure \ref{Fig:Scales}, a). Hence, in the following, the photon gas is assumed to be always in a thermal steady state and the beginning of the experiment refers to right after the thermalisation. This justifies to interpret the quantum mechanical expectation value $N_{\bm l}(t) = \langle \hat a^\dagger_{\bm l}(t) \hat a_{\bm l}(t) \rangle$ as the Bose-Einstein distribution
\begin{align}
N_{\bm l}(t) = \left\{e^{\beta\left[E_{\bm l}(t)-\mu(t)\right]}-1\right\}^{-1}\,.
\end{align}
Here, $E_{\bm l}(t)$ denotes the instantaneous eigenenergies of the Hamiltonian matrix \eqref{eq:Hmat}, $\beta=1/(k_\text{B}T)$ is the inverse temperature, and $\mu(t)$ stands for the instantaneous chemical potential, which is fixed by the conserved total particle number $N=\sum_{\bm l} N_{\bm l}(t)$. Moreover, in the thermal steady state the density \eqref{eq:density} appearing in \eqref{eq:Ham2} takes the form
\begin{align}
    n(\vec x, t) = \sum_{\bm l} N_{\bm l}(t) |\psi_{\bm l}(\vec x)|^2\,,
\end{align}
such that the interaction matrix \eqref{eq:Fmat1} is finally given by
\begin{align}\label{eq:Fmat}
F_{\bm n, \bm{n'}} = \sum_{\bm l} N_{\bm l}(0) \int d^2x~\psi_{\bm n}^* (\vec x) |\psi_{\bm l}(\vec x)|^2 \psi_{\bm{n'}}(\vec x)\,.
\end{align}
Here only the photon occupation from the beginning of the experiment appears due to the temporal retardation of the interaction.\\
Note that this treatment can be seen as a Hartree-Fock analogue for the thermo-optic interaction. Like the standard Hartree-Fock known from atomic and solid-state physics, our method approximates a thermalised, interacting system by transforming to a new eigenbasis, such that the interaction effects are incorporated for calculating physical quantities. Whereas, the standard Hartree-Fock theory for an instantaneous interaction leads to a self-consistency problem, which needs to be solved iteratively, the temporal retardation of the thermo-optic interaction, see figure \ref{Fig:Scales} b) and Hamiltonian \eqref{eq:Ham2}, intrinsically avoids this problem. The latter bears the main physical difference between our adaption of the Hartree-Fock method for a retarded interaction and the established Hartree-Fock method for an instantaneous interaction.
%
\subsection{Potentials}
The formulation of the second-quantised Hamiltonian matrix \eqref{eq:Hmat} is generally valid for any trapping potential. For the purpose of illustration the next section focuses in detail on two concrete trapping potentials. On the one hand, an isotropic harmonic potential of the form
\begin{align}\label{eq:Vho}
V_\text{Ho} = \frac{\hbar\Omega}{2}\,\frac{x^2+y^2}{l^2}
\end{align}
is considered, with the trapping frequency $\Omega$ and the oscillator length $l = \sqrt{\hbar/m\Omega}$. On the other hand, also the box potential 
\begin{align}\label{eq:Vbox}
	V_\text{Box} = \begin{cases} 0, & 0\leq x\leq L, \hspace*{5mm}0\leq y\leq L\\
	\infty, & \text{elsewhere}	\end{cases}\,,
\end{align}
is analysed, where $L$ denotes the width of the box in both directions.
\section{First-Order Perturbation Theory}
\label{sec:Perturbation}
\begin{figure}[t]
	\centering
	\includegraphics[width=\linewidth]{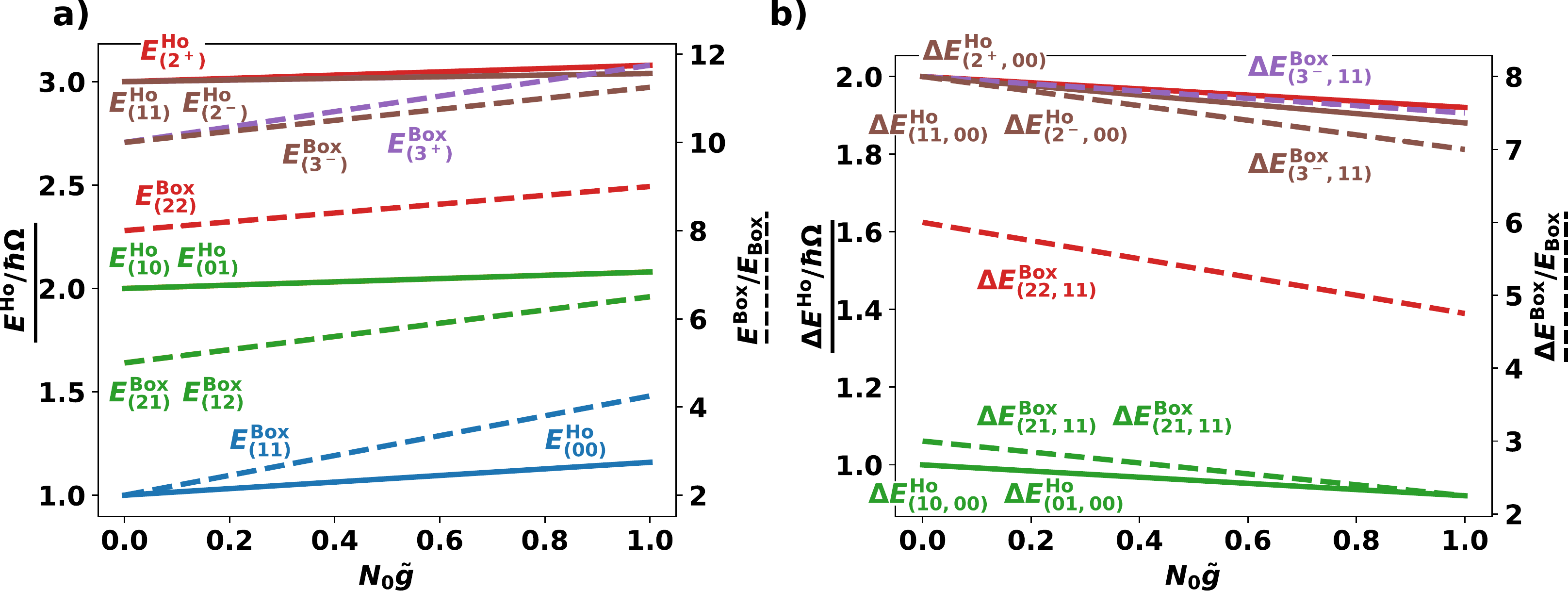}
	\caption{First-order perturbative calculation of eigenenergies. \textbf{a)} Absolute eigenenergies in the form of \eqref{eq:energy_pert} as a function of the interaction strength.  \textbf{b)} Corresponding energy differences to the ground-state energies $E^\text{Ho}_{(00)}$ and $E^\text{Box}_{(11)}$, respectively. In both pictures, the solid lines and the left $y$-axis refer to the harmonic potential \eqref{eq:Vho}, whereas the dashed lines and the right $y$-axis correspond to the box potential \eqref{eq:Vbox}. The different mode indices are summarised in table \ref{Tab:Epert}.}
	\label{Fig:Diffs_pert}
\end{figure}

Since the photon-photon interaction is small, calculating the first-order in Rayleigh-Schrödinger perturbation theory offers initial insights into the physics contained in the Hamiltonian matrix \eqref{eq:Hmat}. Note that first-order perturbation theory neglects the interaction between different energy subspaces in the interaction matrix \eqref{eq:Fmat}.\\
Although reaching a condensate fraction of more than 50 \% is a hard task in current phBEC experiments, this section concentrates on the theoretical analysis of the deep condensate limit, where the ground-state occupation number $N_{\bm 0}$ coincides approximately with the total particle number $N$, i.e., $N_{\bm 0} \approx N$. Hence, the interaction matrix \eqref{eq:Fmat} reduces to
\begin{align}\label{eq:Fsimp}
F_{\bm \alpha, \bm{\beta}} \approx N_{\bm 0}(0)\int d^2x~\psi_{\bm \alpha}(\vec x)|\psi_{\bm 0}(\vec x)|^2\psi_{\bm{\beta}}^*(\vec x) \,,
\end{align}
where, the indices $\bm \alpha$, $\bm{\beta}$ belong to the same energy subspace, such that $E_{\bm \alpha}(0) = E_{\bm{\beta}}(0)$. Consequently, the instantaneous eigenenergies $E_{\bm n}(t)$ have the approximate form
\begin{align}\label{eq:energy_pert}
E_{\bm n}(t) \approx E_{\bm n}(0) + g(t)N_{\bm 0} {\mathcal E}_{\bm n}^{(1)}\,,
\end{align}
where $\mathcal{E}_{\bm n}^{(1)}$ denotes the first-order correction of the $\bm n$th eigenenergy. \ref{App:Perturbation Theory} summarises the respective details of calculating these corrections, and table \ref{Tab:Epert} lists the  corresponding results.\\
The perturbative calculation aims at the energy differences $\Delta E_{{\bm n}, {\bm{n'}}}(t) = E_{\bm n}(t)-E_{\bm{n'}}(t)$ between two different modes. In the considered precision they are given by
\begin{align} \label{eq:DE}
\Delta E_{{\bm n}, {\bm{n'} }}(t) \approx E_{\bm n}(0)-E_{\bm{n'}}(0) + g(t)N_{\bm 0} \left[{\mathcal E}_{\bm n}^{(1)} - {\mathcal E}_{\bm{n'}}^{(1)}\right]\,.
\end{align}
Figure \ref{Fig:Diffs_pert}, a) shows
\begin{table}[t]
	\centering
	\caption{First-order correction to energy eigenvalues \eqref{eq:energy_pert} for \textbf{a)} the harmonic potential \eqref{eq:Vho} and \textbf{b)} the box potential \eqref{eq:Vbox}. The multi-index $\bm n$ takes here the form $\bm n = (n_x\, n_y)$ and the square brackets denote the mode hybridisation due to the thermo-optic interaction.}
	\begin{minipage}[t]{.45\linewidth}
		\textbf{a)}
	\begin{tabular}{|c|c|}
		\hline
		$\bm n$ & $l^2 {\cal E}_{\bm n}^{(1)}$\\
		\hline
		(00) & $1 / (2\pi)$\\
		(10) & $1 / (4\pi)$\\
		(01) & $1 / (4\pi)$\\
		$2^+ = [(20) + (02)]/\sqrt{2}$ & $1 / (4\pi) $\\
		(11) & $1 / (8\pi)$\\
		$2^- =[(20) - (02)]/\sqrt{2}$ & $1 / (8\pi) $\\
		\hline
	\end{tabular}
	\end{minipage}
	\begin{minipage}[t]{.45\linewidth}
		\textbf{b)}
	\begin{tabular}{|c|c|}
		\hline
		$\bm n$ & $L^2 {\cal E}_{\bm n}^{(1)}$\\
		\hline
		(11) & 9/4\\
		(21) & 3/2\\
		(12) & 3/2\\
		(22) & 1\\
		$3^+ = [(31) + (13)]/\sqrt{2}$ & 7/4\\
		$3^- =[(31) - (13)]/\sqrt{2}$ & 5/4\\
		\hline
	\end{tabular}
	\end{minipage}
	\label{Tab:Epert}
\end{table}
the corresponding eigenenergies  up to the second excited states. They are plotted versus the dimensionless interaction strength $\tilde g(t) = m g(t)/\hbar^2$ \cite{Bloch2008}, which depends linearly on time $t$ according to \eqref{eq:g}. With this choice of the scaling, the results presented here are broadly valid for different experimental settings and do not depend on the material specific parameters $\gamma$ and $B$. In addition, the chosen maximum value of 1 for $N_{\bm 0}\tilde g(t)$ corresponds to existing measurements \cite{Klaers2010}. As a consequence of the repulsive thermo-optic interaction, the eigenenergies are generically shifted to larger values. Depending on the mode symmetry, the interaction also lifts some degeneracies, for details see table \ref{Tab:Epert}. Moreover, the thermo-optic interaction influences the ground state most, and is less relevant for higher excited states. Since the effective photon-photon interaction causes these energy shifts, its strength can be extracted from them. Spectroscopic measurements offer experimental access to the energy differences between these modes, as depicted in figure \ref{Fig:Diffs_pert}, b).\\
A comparison
of the results from the different potentials shows that in the case of the box potential the interaction effects are more dominant, which is due to the much stronger confinement of the photon gas. In the harmonic potential the width of the excited states increases since for higher excitations the confinement effectively weakens. This leads to smaller interaction matrix elements \eqref{eq:Fsimp} and, hence, to a smaller effective interaction strength. In case of the box potential, however, this is not possible due to the Dirichlet boundary conditions upon the condensate wave function.\\
The result \eqref{eq:DE} of the first-order perturbation theory yields analytical formulas, which allow determining the effective photon-photon interaction. In the harmonic case, the corresponding energy difference $\Delta E_{(10,00)}^\text{Ho}$ between the first excited and the ground state yields for the effective photon-photon interaction strength the formula
\begin{align}\label{eq:g_det}
\tilde g_\text{Ho}(t) = \frac{4\pi}{N_{\bm 0}}\left[1-\frac{\Delta E_{(10,00)}^\text{Ho}(t)}{\hbar\Omega}\right]\,,
\end{align}
while for the box potential \eqref{eq:Vbox} this correspondingly amounts to
\begin{align}\label{eq:box_eq}
\tilde g_\text{Box}(t) = \frac{2\pi^2}{3N_{\bm 0}}\left[3-\frac{\Delta E_{(21,11)}^\text{Box}(t)}{E_\text{Box}}\right]\,,
\end{align}
with the 1D ground-state energy $E_\text{Box} = \pi^2\hbar^2/(2mL^2)$. Similar formulas can be derived for other combinations of the energy eigenstates.\\
Thus, equations \eqref{eq:g_det} and \eqref{eq:box_eq} offer the prospect for spectroscopically determining the effective photon-photon interaction strength with a higher precision than in previous measurements, which were based on detecting the condensate width \cite{Klaers2010}. For instance, interfering two cavity eigenmodes results in a beating signal, the frequency of which corresponds to the energy difference \eqref{eq:DE} of the involved modes.

\section{Summary and Outlook}
The theory presented in this work is crucial for understanding and precisely quantifying the effective photon-photon interaction strength in current and future phBEC experiments. The second-quantised Hamiltonian \eqref{eq:Ham2} represents the backbone of this theory. It describes the impact of the thermo-optic photon-photon interaction emerging during a single pump pulse. The experimental timescales allow treating the thermo-optic photon-photon interaction adiabatically, yielding a matrix formulation \eqref{eq:Hmat} of the Hamiltonian \eqref{eq:Ham2}. With this, it is possible to predict the shifts of the photon eigenenergies, which allow a precise spectroscopic measurement of the emerging photon-photon interaction.\\
The Hamiltonian matrix \eqref{eq:Hmat} also contains information about the excited states, which permits calculating the impact of the thermal cloud on the measurement. However, this impact is only relevant in case of an increased effective photon-photon interaction, as it may occur at the dimensional crossover \cite{Stein2022b}. Due to the increased photon-photon interaction, a perturbative approach like the one used in section \ref{sec:Perturbation} is no longer valid. Therefore, in such a situation, Exact Diagonalisation turns out to be a necessary tool for analysing the Hamiltonian matrix \eqref{eq:Hmat} \cite{Stein2022d}.
\ack
We thank Antun Bala\v{z}, Georg von Freymann, Milan Radonji\'c, Julian Schulz, Kirankumar Karkihalli Umesh, and Frank Vewinger for insightful discussions. ES and AP acknowledge financial support by the Deutsche Forschungsgemeinschaft (DFG, German Research Foundation) via the Collaborative Research Center SFB/TR185 (Project No. 277625399).
\appendix

\section{Effective Steady State of Temperature}
\label{app:steady_temp}
The thermo-optic photon-photon interaction highly depends on the temperature diffusion inside the microcavity. In particular, the strength of the thermo-optic interaction is proportional to the temperature accumulated during a single experiment, as figure \ref{Fig:Scales}, a) depicts. This is especially important, as the maxi\-mum condensate lifetime is limited by the dye bleaching, as outlined in the introduction. Therefore, this appendix deals with the quasi steady-state of the temperature amplitude during a whole experimental sequence, as it is depicted in figure \ref{Fig:Scales} c). As the temperature decay time $\tau$ is much larger than the condensate lifetime $t_\text{exp}$, a $\delta$-function peak models the temperature gain $\Delta T$ during a single experiment. For the detailed timescales, consider figure \ref{Fig:Scales} a). The temperature gain is connected to the heating rate $B$, the condensate lifetime $t_\text{exp}$ and the photon number $N$ in a single pump pulse via
\begin{align}\label{eq:app-temp0}
	\Delta T = t_\text{exp} B N\,.
\end{align}
In this appendix, $\Delta T$ is assumed to be equal throughout an experimental cycle. Hence, the equation
\begin{align}\label{eq:app-temp1}
	\dot T = -\frac{1}{\tau}T + \Delta T \sum_{n=1}^{M} \delta(t-\Delta t n)
\end{align}
describes the temperature amplitude during a whole sequence of $M$ experiments, each separated by the time $\Delta t$. A Dirac comb models the sequence of experiments heating the cavity. The first step for working out the solution of \eqref{eq:app-temp1} is to consider a single pulse at $t = \Delta t$. Integrating equation \eqref{eq:app-temp1} from $\Delta t-\epsilon$ to $\Delta t + \epsilon$ and taking the limit $\epsilon \rightarrow 0$ yields the temperature gain $\Delta T$ during this pulse. For $t>\Delta t$, equation \eqref{eq:app-temp1} reduces to a simple decay equation. Hence, for a single pulse at $t=\Delta t$ the temperature amounts to
\begin{align}
	T = \begin{cases} 0&, 0\leq t<\Delta t \\ \Delta T e^{-(t-\Delta t)/\tau} &, t\geq\Delta t \end{cases}\,.
\end{align}
Combining now two $\delta$-function peaks leads with the same reasoning to
\begin{align}
	T = \begin{cases} 0&, 0\leq t<\Delta t \\ \Delta T e^{-(t-\Delta t)/\tau} &, \Delta t\leq t<2\Delta t \\ \Delta T \left(1+e^{-\Delta t/\tau}\right) e^{-(t-2\Delta t)/\tau} &, t\geq 2\Delta t\end{cases}\,.
\end{align}
Therefore, the temperature amplitude $T_M$ after $M$ experiments, i.e., $T_M = T(\Delta t M)$ takes the recursive form
\begin{align}\label{eq:app-temp2}
	T_M = \Delta T + T_{M-1}e^{-\Delta t/\tau}\,,
\end{align}
with $T_1 = \Delta T$ and formally defining $T_0=0$. Bringing \eqref{eq:app-temp2} into explicit form and solving the resulting geometric series yields for the total temperature
\begin{align}\label{eq:app-temp3}
	T_M = \Delta T \frac{1-e^{-M \Delta t/\tau}}{1-e^{-\Delta t/\tau}}\,.
\end{align}
The maximum temperature of \eqref{eq:app-temp3} corresponds


to the limit $M\rightarrow\infty$ and is given by
\begin{align}\label{eq:app-temp4}
	T_\text{max} = \frac{\Delta T}{1-e^{-\Delta t/\tau}}\,.
\end{align}
Equation \eqref{eq:app-temp4} shows how the maximal temperature depends on $\Delta T$ and the ratio $\Delta t/\tau$.
\\
In the case of $\Delta t\gg \tau$ subsequent experiments do not influence each other, as here $T_\text{max} = \Delta T$, i.e., the maximally reached temperature is just the one from a single experiment. On the other hand, realistic experimental settings \cite{Klaers2011} amount to $\tau \sim \SI{1}{\s}$ and $\Delta t \approx \SI{8}{\milli\s}$, realising the opposite case $\tau \gg \Delta t$. With \eqref{eq:app-temp0} the maximal temperature \eqref{eq:app-temp4} acquires the form
\begin{align}\label{eq:app-temp5}
	T_\text{max} = \sigma \tau B N\,,
\end{align}
where $\sigma = t_\text{exp}/\Delta t$ is the so-called duty cycle of the experiment, which has in most experiments the value $\sigma = 1/\num{16000}$ due to the condensate lifetime $t_\text{exp}=\SI{500}{\nano\s}$. Thus, the larger the duty cycle, i.e., the condensate lifetime, the larger the interaction.

\section{Perturbation Theory}
\label{App:Perturbation Theory}
This appendix deals with the calculation of perturbative results used in section \ref{sec:Perturbation}.
\subsection{Harmonic Oscillator}
The eigenenergies of the harmonic potential \eqref{eq:Vho} read 
\begin{align}\label{eq:E0ho}
	E_{\bm n}(0) = \hbar\Omega\left( n_x + n_y + 1  \right)
\end{align}
and the corresponding eigenfunctions are the Gauß-Hermite functions
\begin{align}\label{eq:Psiho}
	\psi_{\bm n}(\vec x) = \sqrt{\frac{1}{2^{n_x+n_y}n_x!n_y!\pi l^2}} \,H_{n_x}\left(\frac{x}{l}\right)H_{n_y}\left(\frac{y}{l}\right)e^{-(x^2+y^2)/(2l^2)}\,,
\end{align}
where $H_n(x)$ denote the Hermite polynomials. The eigenfunctions \eqref{eq:Psiho} give rise to the simplified interaction matrix \eqref{eq:Fsimp} up to the third energy subspace
\begin{align}\label{eq:FHo}
	F_\text{Ho} = \frac{N_{(00)}}{\pi l^2} \begin{pmatrix}
		1/2 & 0 & 0 & -1/(4\sqrt{2}) &0&-1/(4\sqrt{2})\\
		0 & 1/4 & 0&0&0&0\\
		0&0& 1/4&0&0&0\\
		-1/(4\sqrt{2}) & 0&0& 3/16 &0& 1/16\\
		0 &0&0&0& 1/8 &0\\
		-1/(4\sqrt{2}) &0&0& 1/16 &0& 3/16
		\end{pmatrix}\, ,
\end{align}
where the modes are ordered with respect to their vectorised mode indices, c.f., table \ref{Tab:Ordering}, a). Calculating the first-order corrections to the energies amounts to neglecting the interaction between different energy subspaces, i.e., neglecting the coupling between ground state and second excited state in \eqref{eq:FHo}. The resulting eigenvalues ${\cal E}_{\bm n}^{(1)}$ are shown in table \ref{Tab:Epert}.
\subsection{Box Potential}
The eigenenergies of the box potential \eqref{eq:Vbox} are given by
\begin{align}
	E_{\bm n}(0) = E_\text{Box}\left(n_x^2+n_y^2\right)\,,
\end{align}
with the 1D ground-state energy $E_\text{Box} = \pi^2\hbar^2/(2mL^2)$ and the eigenfunctions read
\begin{align}
	\psi_{\bm n} (\vec x) = \frac{2}{L}\sin\left(\frac{n_x\pi x}{L}\right)\sin\left(\frac{n_y\pi y}{L}\right)\,.	
\end{align}
This yields for the simplified interaction matrix \eqref{eq:Fsimp}
%

\begin{align}
	F_\text{Box} = \frac{N_{(11)}}{L^2} \begin{pmatrix}
		9/4 & 0& 0&0&-3/4&-3/4\\
		0&3/2 &0&0&0&0\\
		0&0&3/2&0&0&0\\
		0&0&0&1&0&0\\
		-3/4&0&0&0&3/2&1/4\\
		-3/4&0&0&0&1/4&3/2
		\end{pmatrix}\,,
\end{align}
which is treated as the corresponding one for the harmonic potential \eqref{eq:FHo}.

\begin{table}
	\centering
	\caption{Index ordering \textbf{a)} for harmonic potential \eqref{eq:Vho} and \textbf{b)} for box potential \eqref{eq:Vbox} in the corresponding Hamiltonian matrix.}
	\begin{minipage}[t]{.45\linewidth}
		\textbf{a)}
	\begin{tabular}{|c|c|}
		\hline
		$\bm n$ & \textbf{Vectorised index}\\
		\hline
		(00) & 1\\
		(10) & 2\\
		(01) & 3\\
		(20) & 4\\
		(11) & 5\\
		(02) & 6\\
		\hline
	\end{tabular}
	\end{minipage}
	\begin{minipage}[t]{.45\linewidth}
		\textbf{b)}
	\begin{tabular}{|c|c|}
		\hline
		$\bm n$ & \textbf{Vectorised index}\\
		\hline
		(11) & 1\\
		(21) & 2\\
		(12) & 3\\
		(22) & 4\\
		(31) & 5\\
		(13) & 6\\
		\hline
	\end{tabular}
	\end{minipage}
	\label{Tab:Ordering}
\end{table}

\section*{References}
\bibliographystyle{unsrt}
\bibliography{refs}

\end{document}